\documentstyle[aps,prb,twocolumn,eqsecnum,epsf,floats]{revtex}

\begin{document}

\draft
\preprint{}

\twocolumn[\hsize\textwidth\columnwidth\hsize\csname @twocolumnfalse\endcsname

\title{Noise in Multiterminal Diffusive Conductors: \\
       Universality, Nonlocality and Exchange Effects}

\author{Eugene V.\ Sukhorukov\cite{Eugene}\cite{inst}
        and Daniel Loss\cite{Daniel}}
\address{Department of Physics and Astronomy,
         University of Basel,\\
         Klingelbergstrasse 82,
         CH--4056 Basel, Switzerland}
\date{\today}
\maketitle
\begin{abstract}
We study noise and transport in multiterminal diffusive conductors. Using a
Boltzmann-Langevin equation approach we reduce the calculation of shot-noise
correlators to the solution of diffusion equations. Within this approach we prove
the universality of shot noise in multiterminal diffusive conductors of arbitrary
shape and dimension for purely elastic scattering as well as for hot electrons.
We show that shot noise in multiterminal conductors is a non-local quantity and
that exchange effects can occur in the absence of quantum phase coherence even at
zero electron temperature. It is also shown that the exchange effect measured in
one contact is always negative -- in agreement with the Pauli principle. We
discuss a new phenomenon in which current noise is induced by thermal transport.
We propose a possible experiment to measure locally the effective noise
temperature. Concrete numbers for shot noise are given that can be tested
experimentally.
\end{abstract}

\vskip2pc]
\narrowtext

\section{Introduction}
\label{introduction}

Shot noise is a nonequilibrium fluctuation of the 
current in mesoscopic conductors caused by random flow
of the charge. It can be thought of as an uncorrelated
Poisson process \cite{Schottky} giving rise to a simple formula
for the spectral density of the shot noise,
$S^c=eI$, where  
$I$ is the current through the conductor
and $e$ is the electron charge.
Being the result of charge quantization, the
shot noise is an interesting and highly nontrivial 
physical phenomenon.\cite{Jongrev} In contrast to the 
thermal fluctuations of the current, the shot noise
provides important information about microscopic
transport properties of the conductors beyond 
the linear response coefficients such as the conductance.
For instance, 
the shot noise serves
as a sensitive tool to study correlations
in conductors:
while shot noise assumes the Poissonian value in the absence of correlations,
it becomes suppressed when correlations set in as e.g.
imposed by the Pauli principle.
\cite{Khlus,Landauer,Lesovik,Yurke,Buttiker1}
In particular, the shot noise is completely suppressed
in ballistic conductors,\cite{Kulik}
and it appears thus only in the presence of a disorder.

\begin{figure}
  \begin{center}
    \leavevmode
\epsfxsize=7.0cm
\epsffile{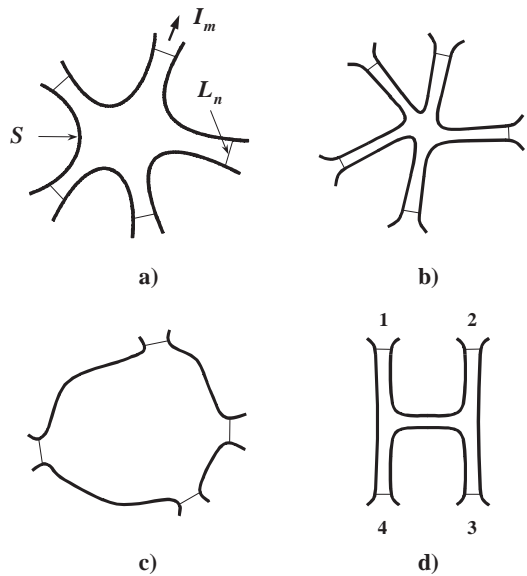}
  \end{center}
\caption{
a) Multiterminal diffusive conductor of arbitrary 2D or 3D shape
and with arbitrary impurity distribution.
There are $N$ leads with metallic contacts of area $L_n$,
and $I_m$ is the m-th outgoing current.
$S$ denotes the remaining surface of the conductor where no current
can pass through. 
b) Conductor of a star  geometry with $N$ long leads
which join each other at a small crossing
region. The resistance of this region is assumed to be
much smaller than the resistance of the leads.
c) Wide	conductor: the contacts are connected through a wide
region, so that the resistance of the conductor comes mainly from 
the regions near the contacts, while the resistance of the wide 
region is negligible.
d) H-shaped conductor with four leads of equal
conductances, $G/4$, connected
by a wire in the middle of conductance $G_0$.} 
\label{noisefig}
\end{figure}

In diffusive mesoscopic
two-terminal conductors where the inelastic scattering lengths
exceed the system
size the shot-noise suppression factor for ``cold'' electrons
(i.e. for vanishing electron temperature)
was predicted
\cite{Beenakker1,Nagaev1,Jong1,Nazarov,Altshuler,Jong}
to be $1/3$.  The suppression of shot noise in diffusive conductors
is now experimentally
confirmed. 
\cite{Liefrink,Steinbach,Schoelkopf,Schoenenberger,Schoenenberger2}
While some derivations 
are based on a scattering matrix approach \cite{Beenakker1,Jong1}
or conventional Green's function technique \cite{Nazarov,Altshuler}
and thus a priori include
quantum phase coherence, no such effects  are included in the
semiclassical Boltzmann-Langevin equation approach,
which nevertheless leads to the same result.\cite{Nagaev1,Jong} 
However,
while in the quantum approach for a two-terminal conductor
the factor $1/3$ was even shown to be universal,\cite{Nazarov}
the semiclassical derivations given so far \cite{Nagaev1,Jongrev}
are restricted to quasi--onedimensional conductors.
Thus, although phase coherence is believed not to be essential for
the suppression	of shot noise,\cite{Shimitzu}
the equivalence of different approaches for calculating
noise in mesoscopic conductors is not evident.
In the regime of hot electrons the noise suppression factor 
was found\cite{Nagaev2,Kozub} to be $\sqrt{3}/4$.
Again, this result, which is based on a Boltzmann-Langevin 
equation approach, is restricted to quasi--onedimensional conductors.
The generalization of these results  
to the case of arbitrary multiterminal conductors is not obvious.

We present here the systematic study of transport and noise 
in multiterminal diffusive conductors. 
This problem has been recently addressed by 
Blanter and B\"{u}ttiker in Ref.\ \onlinecite{Blanter}, where they use
the scattering matrix formulation followed by an impurity averaging
procedure. Having the advantage of including quantum phase coherence,
this approach is somewhat cumbersome to generalize to an arbitrary
geometry and arbitrary disorder.
In contrast to this, 
our approach is based on semiclassical Boltzmann-Langevin equation, 
which greatly 
simplifies the calculations. 

We consider a multiterminal mesoscopic
diffusive conductor (see Fig.\ \ref{noisefig}a) connected to
an arbitrary number $N$ of perfect metallic reservoirs at the
contact surfaces
$L_n$, $n=1,\ldots ,N$, where the voltages $V_n$ or outgoing currents $
I_n$
are measured. The reservoirs are maintained at equilibrium
and have in general different lattice temperatures $T_n$.
Unless specified otherwise the conductor has an arbitrary
3D or 2D geometry with an arbitrary disorder distribution.
Our goal is to calculate the multiterminal spectral densities
of current fluctuations $\delta I_n(t)$
at zero frequency, $\omega =0$,
\begin{equation}
S^c_{nm}=\int\limits_{-\infty}^{\infty}dt\langle\delta I_n(t)\delta
I_m(0)\rangle,
\label{spectr}
\end{equation}
where the brackets $\langle\ldots\rangle$ indicate an  ensemble average.
We consider the effects of purely elastic
scattering and  those of  energy relaxation due to
electron-electron and electron-phonon scattering on the
same basis.

Starting our analysis with a brief summary of the 
Boltzmann-Langevin kinetic equation approach,\cite{Kadomtsev,Kogan}
we then apply the standard diffusion approximation and
reduce the problem of evaluating Eq.\ \ref{spectr}
to the solution of a diffusion equation.
First, we solve the diffusion equation for the distribution function 
to obtain the multiterminal conductance matrix and energy transport
coefficients in terms of well defined 
``characteristic potentials''.\cite{Buttiker3}
We formulate the Wiedemann-Franz law for the case 
of a multiterminal conductor.
Then we turn to the calculation of the noise spectrum.
We derive the exact general formula (\ref{MSD}) for the multiterminal
spectral density of the noise, which together with Eqs.\ 
(\ref{temp},\ref{temp3},\ref{Pi2},\ref{MSD4a}) is the central result 
of our paper.
Using this formula we demonstrate
that the shot-noise suppression
factor of $1/3$ is  {\it universal}
also in the semiclassical Boltzmann-Langevin approach,
in the sense that
it holds for a
multiterminal diffusive conductor of  arbitrary shape, 
electron spectrum and disorder distribution.
We first prove this for cold electrons and then for the case of
hot electrons where the suppression factor is $\sqrt{3}/4$.
Thereby we extend previous semiclassical
investigations\cite{Nagaev2,Kozub} for
two-terminal conductors
to an arbitrary multiterminal geometry.
This allows us then to compare our semiclassical approach
with the scattering matrix approach
for multiterminal conductors,\cite{Buttiker1,Buttiker2,Martin1}
in particular with some explicit results recently obtained
for diffusive conductors.\cite{Blanter}
The universality of shot noise proven here gives further
support to the suggestion
\cite{Jong2} that phase coherence is not essential for the
suppression of shot noise in diffusive conductors.\cite{Landauer2}

Another remarkable property of shot noise in mesoscopic conductors is the
exchange effect introduced by B\"uttiker.\cite{Buttiker2} Although this effect
is generally believed to be phase-sensitive, we will show that this need not be
so. Indeed, for the particular case of an H-shaped conductor (see Fig.\
\ref{noisefig}d) we show that exchange effects can be of the same order as the
shot noise itself even in the framework of the semiclassical Boltzmann approach.
We prove that while the exchange effect measured in different contacts
(cross-correlations) can change the sign, it is always negative when measured in
the same contact (auto-correlations). Thus, the auto-correlations are always
suppressed, in agreement with the Pauli principle. Formally, these exchange
effects are shown to come from a non-linear dependence on the local distribution
function. Similarly we show that the same non-linearities are responsible for
non-local effects such as the suppression of shot noise by open leads even at
zero electron temperature.

Finally, we discuss a new phenomenon, namely the current noise in multiterminal
diffusive conductors induced by thermal transport. We consider the cases of hot
and cold electrons and prove the universality of noise in the presence of
thermal transport. We also propose a possible experiment which would allow one to
measure locally the effective noise temperature. Throughout the paper we
illustrate the general formalism introduced here by concrete numbers for various
conductor shapes that are of direct experimental interest.
We note that some of results of present paper has been published 
in Ref.\ \onlinecite{sukhor} in less general form. Here we present the
details of the derivation of these results and generalize them to a
finite temperature and an arbitrary electron spectrum (band structure).

\section{Boltzmann-Langevin equation: diffusive regime}
\label{bl}

To calculate the spectral density of current fluctuations
we use the Boltzmann-Langevin kinetic
equation \cite{Kadomtsev,Kogan}
for the fluctuating distribution function $F({\bf p},{\bf r},t)=f(
{\bf p},{\bf r})+\delta f({\bf p},{\bf r},t)$,
which depends on the momentum ${\bf p}$, position ${\bf r}$, and time $
t$,
\begin{equation}
\left(\partial_t+{\bf v}\!\cdot\!\partial_{{\bf r}}+e{\bf E}\!\cdot\!
\partial_{{\bf p}}\right)F-I[F]-I_{im}[F]=\delta F^s,
\label{kineq}
\end{equation}
where ${\bf E}({\bf r},t)={\bf E}({\bf r})+\delta {\bf E}({\bf r},t)$
is the fluctuating electric field, ${\bf v}
=\nabla_{{\bf p}}\varepsilon$ is the velocity
of the electron and $\varepsilon$ is its kinetic energy.
$I[F]=I_{ee}[F]+I_{e-ph}[F]$ contains
the electron-electron and electron-phonon collision integrals, respectively
(we do not need to specify them here),
and $I_{im}[F]$ is the impurity collision integral,
\begin{eqnarray}
\lefteqn{
I_{im}[F]=\sum_{{\bf p}^{\prime}}\left(J_{{\bf p}^{\prime}{\bf p}}
-J_{{\bf p}{\bf p}^{\prime}}\right),
}\nonumber \\
& &
J_{{\bf p}{\bf p}^{\prime}}
({\bf r},t)=W_{{\bf p}{\bf p}^{\prime}}({\bf r})F({\bf p},{\bf r},
t)[1-F({\bf p}^{\prime},{\bf r},t)],
\label{coll}
\end{eqnarray}
where the elastic scattering rate from ${\bf p}$ into ${\bf p}^{\prime}$,
$W_{{\bf p}{\bf p}^{\prime}}({\bf r})$,
depends on the position ${\bf r}$ in the case
of disorder considered here.

The Langevin source of fluctuations $\delta F^s({\bf p},{\bf r},t)$
is induced
by the random (stochastic) process of the electron scattering  
which is also responsible
for the momentum relaxation of the electron gas.
On the other hand,  electron-electron scattering
conserves total momentum of the electron gas and therefore does not
contribute to $\delta F^s$. 
Furthermore, we neglect the momentum relaxation due to
electron-phonon scattering 
and electron-electron Umklapp process, assuming that they are weak
compared to the scattering by impurities in diffusive
conductors (phonon induced shot noise in ballistic wires has been
studied in Ref.\ \onlinecite{Gurevich}).
In other words, we assume that the collision integrals $I_{ee}[F]$ and
$I_{e-ph}[F]$ describe only energy relaxation process in the electron
gas, but it is only impurity scattering which gives rise to
momentum relaxation and to the shot noise in 
diffusive conductors.

To describe the fluctuations $\delta F^s$ we make use of the
Langevin formulation introduced by
Kogan and Shul'man (Ref.\ \onlinecite{Kogan}).
In this approach
there are two contributions to the fluctuations
of the impurity collision integral. 
First, there is the contribution $I_{im}[\delta
f]$
due to the fluctuations of the distribution function, which has
already been  included in  Eq.\ (\ref{coll}).
The second contribution, $\delta I_{im}[f]$, 
stems from the random character
of the electron scattering, which is 
the extra source
of fluctuation $\delta F^s$ occurring on the {\it rhs\/} of 
Eq.\ (\ref{kineq}), i.e.,
\begin{equation}
\delta F^s=\sum_{{\bf p}^{\prime}}\left(\delta J_{{\bf p}^{\prime}
{\bf p}}-\delta J_{{\bf p}{\bf p}^{\prime}}\right),
\label{source}
\end{equation}
where the random variables $\delta J_{{\bf p}{\bf p}^{\prime}}$
are intrinsic fluctuations of the incoming and
outgoing fluxes $J_{{\bf p}{\bf p}^{\prime}}$.

Assuming now
that the flow of electrons, say, from state ${\bf p}$
to state ${\bf p}^{\prime}$ is described by a Poisson process
we can write\cite{Kogan}
\begin{eqnarray}
\lefteqn{
\left<\delta J_{{\bf p}{\bf p}^{\prime}}({\bf r},t)\delta J_{{\bf p}_
1{\bf p}^{\prime}_1}({\bf r}_1,t_1)\right>
\frac{}{}}\ \ \ \ \ \ \nonumber \\
& &
=\delta (t-t_1)\delta ({\bf r}
-{\bf r}_1)\delta_{{\bf p}{\bf p}_1}\delta_{{\bf p}^{\prime}{\bf p}_
1^{\prime}}\left<J_{{\bf p}{\bf p}^{\prime}}({\bf r},t)\right>,
\label{correlator}
\end{eqnarray}
where
\begin{equation}
\left<J_{{\bf p}{\bf p}^{\prime}}({\bf r},t)\right>=W_{{\bf p}{\bf p}^{
\prime}}({\bf r})f({\bf p},{\bf r})[1-f({\bf p}^{\prime},{\bf r})].
\label{flux}
\end{equation}
Using the preceding two equations together with  Eq.\ (\ref{source}),
we obtain the correlator of the Langevin sources,
\begin{eqnarray}
\lefteqn{
\left<\delta F^s({\bf p},{\bf r},t)
\delta F^s({\bf p}^{\prime},{\bf r}^{\prime},t^{\prime})\right>
=\delta (t-t^{\prime})\delta ({\bf r}-{\bf r}^{\prime})
\frac{}{}
}\nonumber \\
& &
\times\sum_{{\bf p}^{\prime\prime}}
(\delta_{{\bf p}{\bf p}^{\prime}}-
\delta_{{\bf p}^{\prime\prime}{\bf p}^{\prime}})
W_{{\bf p}{\bf p}^{\prime\prime}}
\left[
f(1-f^{\prime\prime})+f^{\prime\prime}(1-f)
\right].
\label{corr}
\end{eqnarray}
with $f^{\prime\prime}\equiv f({\bf p}^{\prime\prime},{\bf r})$,
and $W_{{\bf p}{\bf p}^{\prime\prime}}=W_{{\bf p}^{\prime\prime}{\bf p}}$.

Next, we consider the {\it lhs\/} of  Eq.\ (\ref{kineq}).
Since we are only interested  in the $\omega =0$ limit 
of the spectral density
(the effect of screening on frequency dependent shot noise in quasi--one
dimensional diffusive conductors has been studied recently 
in Refs.\ \onlinecite{Naveh} and \onlinecite{Nagaev3}),
we may drop the first term $\partial F/\partial t$ in Eq.\ (\ref{kineq}).
The term $e{\bf E}\!\cdot\!\partial_{{\bf p}}F$ can be rewritten as follows:
$e{\bf E}\!\cdot\!\partial_{{\bf p}_F}F+e{\bf v}\!\cdot\!{\bf E}\partial_{
\varepsilon}F$, where ${\bf p}_F$ is the momentum at the
Fermi surface. From this we see that
the electric field ${\bf E}$ induced by an applied voltage plays
a twofold role: it effects  the trajectories and changes
the energy of electrons. The first effect, $e{\bf E}\!\cdot\!\partial_{
{\bf p}_F}F\sim eE/p_F$,
is weak compared to ${\bf v}\!\cdot\partial_{{\bf r}}F\sim v_F/L$ ($
L$ is the size of the conductor)
and gives  contribution of order $eV/\varepsilon_F$, which can be
neglected.\cite{estimate} 
The second effect
can be taken into account by the replacement $\varepsilon\to\varepsilon
-eV({\bf r},t)$ in the
argument of the distribution function $F$, so that $\varepsilon$ now
is the total (kinetic $+$ potential) energy of the electron.
Then, the two terms ${\bf v}\!\cdot\!\partial_{{\bf r}}F+e{\bf E}\!\cdot\!
\partial_{{\bf p}}F$ in Eq.\ (\ref{kineq})
can be replaced by the total derivative ${\bf v}\!\cdot\!\nabla F$.

In a next step we apply the standard diffusion approximation 
to the kinetic equation
\cite{Landau} where
the distribution function is split into two parts,
\begin{equation}
F({\bf p},{\bf r},t)=F_0(\varepsilon ,{\bf r},t)+{\bf l}({\bf p}_F,
{\bf r})\!\cdot\!{\bf F}_1(\varepsilon ,{\bf r},t),
\label{approx}
\end{equation}
where the vector ${\bf l}$ obeys the equation,
\begin{equation}
\sum_{{\bf p}^{\prime}}W_{{\bf p}{\bf p}^{\prime}}({\bf r})
[{\bf l}({\bf p}_F,{\bf r})
-{\bf l}({\bf p}^{\prime}_F,{\bf r})]={\bf v}\, .
\label{eqnl}
\end{equation}
The choice of the distribution function $F$ in the form (\ref{approx})
is dictated by the fact that the impurity collision integral
$I_{im}[F]$ does not affect the energy dependence of the distribution 
function.
Inserting this ansatz into Eq.\ (\ref{kineq})  and averaging subsequently 
over the momentum first 
weighted with one and then with ${\bf l}$, we arrive
at 
\begin{equation}
\nabla\!\cdot\!\hat {D}{\bf F}_1-
\overline{I[F]}=0,\label{kineq2a}
\end{equation}
\begin{equation}
\hat{D}(\nabla F_0+{\bf F}_1)=\overline {
{\bf l}\delta F^s}.
\label{kineq2b}
\end{equation}
Here the overbar means averaging over ${\bf p}_F$ at the Fermi surface
inside the Brillouin zone,
$\overline{\left(\ldots\right)}=
\int d{\bf p}_Fv_F^{-1}\left(\ldots\right)/\int d{\bf p}_Fv_F^{-1}$,
and we introduced the diffusion tensor,
\begin{equation}
\hat {D}({\bf r})\equiv D_{\alpha\beta}({\bf r})=\overline {v_{\alpha}
l_{\beta}({\bf p}_F,{\bf r})}.
\label{tensor}
\end{equation}
We also
used  $\overline {\delta F^s}=0$, which follows from Eq.\ (\ref{corr})
and which reflects the
conservation of the number of electrons in the scattering process.

Using the distribution function (\ref{approx}) we can calculate
the current density  ${\bf j}+\delta {\bf j}=e
\nu_F\hat {D}\int d\varepsilon {\bf F}_1$
and due to charge neutrality 
(neglecting accumulation of charge) we get the potential,
$eV+e\delta V=\int_{\varepsilon_
c}^{\infty}d\varepsilon F_0$,
where $\varepsilon_c$ is a constant energy near the 
Fermi level and chosen so that $F|_{\varepsilon_c}=1$,
and $\nu_F=\int d{\bf p}_Fv_F^{-1}$ 
is the density of states at the Fermi level. Upon integration
of Eqs.\ (\ref{kineq2a}) and (\ref{kineq2b})
over the energy $\varepsilon$ the collision integrals
vanish and we arrive at the diffusion equations for the potential and
density of current, resp.,
\begin{equation}
\nabla\!\cdot\!\hat{\sigma}\nabla V=0,\quad {\bf j}=-\hat{\sigma}
\nabla V,\label{diffa}
\end{equation}
\begin{equation}
\delta {\bf j}+\hat{\sigma}\nabla\delta V=
\delta {\bf j}^s,\quad\nabla\!\cdot\!\delta {\bf j}=0,
\label{diffb}
\end{equation}
where the conductivity tensor $\hat{\sigma }({\bf r})=e^2\nu_F\hat {
D}({\bf r})$ depends in general on the
position ${\bf r}$, and $\delta {\bf j}^s=e\nu_F\int
d\varepsilon\overline {
{\bf l}\delta F^s}$
is the Langevin
source of fluctuations of the current density. After 
integrating  over $\varepsilon$ in Eq.\ (\ref{corr}) and averaging over
${\bf p}$ (at the Fermi surface) we use then Eqs.\ (\ref{eqnl})
and (\ref{tensor}) to
obtain the correlation function of
the Langevin sources
\begin{eqnarray}
\langle\delta j^s_{\alpha}({\bf r},t)\delta j^s_{\beta}({\bf r}^{
\prime},t^{\prime})\rangle =\delta (t-t^{\prime})\delta ({\bf r}-{\bf r}^{
\prime})\sigma_{\alpha\beta}({\bf r})\Pi ({\bf r}),\nonumber\\\Pi
({\bf r})=2\int d\varepsilon f_0(\varepsilon ,{\bf r})[1-f_0(\varepsilon
,{\bf r})],
\label{corr2}
\end{eqnarray}
where $f_0$ is symmetric part of the average distribution function
$f=f_0+{\bf l}\!\cdot\!{\bf f}_1$.

The physical interpretation of 
Eq.\ (\ref{corr2})
is now transparent: the function $\Pi$ describes the local broadening
of the distribution function and can be thought of as
an effective (noise) temperature. Then we see that
the  correlator (\ref{corr2})
takes an equilibrium-like form
of the fluctuation-dissipation theorem. This is a direct
consequence of our diffusion approximation.
In the diffusive regime all microscopic details of the
transport and fluctuation mechanisms are hidden in the same
conductivity matrix, which appear in the correlator of the fluctuation
sources (\ref{corr2}) as well as in the diffusion equations
(\ref{diffa}, \ref{diffb}).
It is this
fact which leads to the universality of shot noise 
that is independent of microscopic mechanisms of the noise.

Next, subtracting the fluctuating part from
Eqs.\ (\ref{kineq2a}) and (\ref{kineq2b})
we get the equations for the average distribution function $f$,
\begin{equation}
\nabla \!\cdot\!\hat\sigma\nabla f_0
+e^2\nu_F\overline{I[f]}=0,\quad
f=f_0-{\bf l}\!\cdot\!\nabla f_0,
\label{diff2}
\end{equation}
which complete the set of coupled equations to be solved.
Now we specify the boundary conditions to be imposed on
Eqs.\ (\ref{diffa}, \ref{diffb}, \ref{diff2}).
First, we
assume that
for a given energy there is no current
through the surface $S$ (see Fig.\ \ref{noisefig}a).
Second, since the contacts with area $
L_n$
are perfect conductors the average potential $V$ and its
fluctuations $\delta V$ are independent of position ${\bf r}$ on
$L_n$. Third,
the contacts are assumed to be in thermal equilibrium
with outside reservoirs.\cite{contactheat}
Then we write the boundary conditions for (\ref{diffa}) and (\ref{diffb}),
respectively, as
\begin{equation}
d{\bf s}\!\cdot\!{\bf j}({\bf r})|_S=0,\quad V({\bf r}
)|_{L_n}=V_n,
\label{bounda}
\end{equation}
\begin{equation}
d{\bf s}\!\cdot\!\delta {\bf j}({\bf r},t)|_S=0,\quad
\delta V({\bf r},t)|_{L_n}=\delta V_n(t),
\label{boundb}
\end{equation}
and for (\ref{diff2}),
\begin{equation}
f_0(\varepsilon ,{\bf r})|_{L_n}\!\! =\! f_{T_n}(\varepsilon\! -\! eV_
n),\;\; d{\bf s}\!\cdot\!\hat\sigma({\bf r})\nabla f_0
(\varepsilon ,{\bf r})|_S\! =\! 0,
\label{bound2}
\end{equation}
where $f_{T_n}(\varepsilon )=\left[1+\exp(\varepsilon /T_n)\right]^{-1}$
is the equilibrium distribution function at
temperature $T_n$, and
$d{\bf s}$ is a vector area element perpendicular to the surface.

Eqs.\ (\ref{diffa}, \ref{diffb}, \ref{diff2}) with the boundary conditions
(\ref{bounda}, \ref{boundb}, \ref{bound2})
are now a complete set of equations. In principle,
these equations can be solved exactly which would allow us
to evaluate $S_{nm}^c$ for an arbitrary multiterminal geometry
of the conductor and for an arbitrary disorder distribution.

\section{Solution of the diffusion equations}
\label{solution}
\subsection{Multiterminal conductance matrix}
\label{conduct}

The multiterminal conductance matrix is defined as follows:
$I_n=\sum_mG_{nm}V_m$ (throughout the paper
the sum over the contacts $m$  runs
from $m=1$ to $m=N$, and we omit the limits for convenience).
To calculate $G_{nm}$ we need to solve
Eqs. (\ref{diffa}) with boundary conditions (\ref{bounda}).
Following B\"uttiker \cite{Buttiker3} we introduce
characteristic potentials $\phi_n({\bf r})$, $n=1,\ldots ,N$, associated
with
the corresponding contacts. These functions
satisfy the diffusion equation and the boundary conditions:
\begin{equation}
\nabla\!\cdot\!\hat\sigma\nabla\phi_n=0, \label{character}
\end{equation}
\begin{equation}
d{\bf s}\!\cdot\!\hat\sigma\nabla\phi_n\left
|\right._S=0,\quad\phi_n|_{L_m}=\delta_{nm},
\label{character2}
\end{equation}
so that they are always positive $\phi_n({\bf r})\geq 0$, $n=1,\ldots
,N$
and obey the sum rule (see Appendix \ref{A}),
\begin{equation}
\sum_n\phi_n({\bf r})=1.
\label{sum}
\end{equation}

The potential $V$ can be expressed in terms of characteristic
potentials
\begin{equation}
V({\bf r})=\sum_n\phi_n({\bf r})V_n
\label{potential}
\end{equation}
to satisfy the diffusion equation
(\ref{diffa}) and boundary conditions (\ref{bounda}).
Then the outgoing current through the $m$-th
contact is  $I_m=\int_{L_m}d{\bf s}\!\cdot\! {\bf j}=-\sum_n\int_{L_m}
d{\bf s}\!\cdot\!\hat\sigma\nabla\phi_n V_n$, and using the
definition of the conductance matrix we get
\begin{equation}
G_{mn}=-\int\limits_{L_m}d{\bf s}\!\cdot\!\hat\sigma\nabla\phi_n.
\label{conductance}
\end{equation}
We note here that the multiplication of the
integrand by $\phi_m$ does not change the integral in the {\em rhs} of this
equation. Moreover, the boundary conditions (\ref{character2}) for
the characteristic potentials allows us to extend the integral to the
entire surface. Doing so and taking into account Eq.\ (\ref{character}),
we then replace the surface integral by
an integral over the volume of the conductor. We are then left
with  another
useful formula for $G_{nm}$,
\begin{equation}
G_{mn}=-\int d{\bf r}\nabla\phi_m\!\cdot\!\hat\sigma\nabla\phi_n.
\label{conductance2}
\end{equation}
From this expression and from the sum rule for $\phi_n$
it immediately follows that $G_{nm}=G_{mn}$, $\sum_nG_{nm}=0$,
and $G_{nn}<0$, as it should be. In Appendix \ref{A} we use a similar
procedure
to prove another quite natural property of the conductance matrix:
$G_{nm}>0$ for $n\neq m$.

\subsection{Energy transport coefficients}
\label{energy}

We have already seen that the local source of noise is defined
by the effective noise temperature $\Pi$ (see Eq.\ (\ref{corr2})),
which describes the broadening of the distribution function.
Another important quantity is given by the energy density
$\Upsilon ({\bf r})$ acquired by the electron gas due to the
broadening of the distribution function (effective heat density).
It is given explicitly by the integral
\begin{equation}
\Upsilon =\nu_F\int\limits_{\varepsilon_c}^{\infty}d\varepsilon\varepsilon\left[
f_0-\theta\left(\varepsilon -eV\right)\right]
=\Lambda -{1\over 2}\nu_F(eV)^2,
\label{edensity}
\end{equation}
where $\theta\left(\varepsilon -eV\right)$ is the local
equilibrium distribution function
at zero temperature, and
$\Lambda ({\bf r})=\nu_F\int_{\varepsilon_c}^{\infty}d\varepsilon\varepsilon
f_0(\varepsilon ,{\bf r})-\nu_F\varepsilon_c^2/2$ is the total energy
density
(up to irrelevant constant).

To calculate $\Upsilon$
we integrate the first of Eqs.\ (\ref{diff2}) over $\varepsilon$ with the
weight
of $\varepsilon$ and use the expression (\ref{edensity}) for $\Lambda$.
Then the electron-electron collision integral vanishes, and
we arrive at the following equation,\cite{comment0}
\begin{equation}
\nabla\!\cdot\hat {D}\nabla\Lambda =\nabla\!\cdot\hat {D}\nabla\Upsilon
+{\bf j}\!\cdot\!{\bf E}=q,
\label{econserv}
\end{equation}
where we introduced the rate of energy relaxation (or absorption) due to
phonons,
$q({\bf r})=-\nu_F\int_{\varepsilon_c}d\varepsilon\varepsilon
\overline{
I_{e-ph}[f]}$.
Eq.\ (\ref{econserv}) expresses energy conservation:
the work done on the system by the electric field, ${\bf j}\!\cdot\!
{\bf E}$,
is equal to 
the energy flux to the lattice, $q$, plus
the heating of the electron gas, $-\nabla\!\cdot\hat {D}
\nabla\Upsilon$.
Integration of Eqs.\ (\ref{bound2}) gives us the boundary
conditions for $\Lambda$,
\begin{eqnarray}
\Lambda |_{L_n}=\Lambda_n=\nu_F\left[{\pi^2\over 6} T_n^2+{1\over
2}(eV_n)^2\right], \nonumber \\
d{\bf s}\!\cdot\!\hat {D}\nabla\Lambda |_S=0.
\label{bound3}
\end{eqnarray}

We assume now that electron-phonon interaction is
weak (the general case is discussed 
in Sec.\ \ref{density}).
Then the energy exchange between the electron gas and the lattice
occurs in the metallic reservoirs far away
from the conductor, and inside the conductor we have $q=0$. 
Eq.\ (\ref{econserv}) for $\Lambda$ with the boundary conditions (\ref{bound3})
can be solved in terms of $\phi_n$: $\Lambda ({\bf r})=\sum_n\phi_
n({\bf r})\Lambda_n$. Substituting
this expression into Eq.\ (\ref{edensity}) and
using  Eq.\ (\ref{potential}) for $V$, we obtain $\Upsilon$,
\begin{equation}
\Upsilon =\nu_F
\sum_{n,m}\phi_n\phi_m
\left[{\pi^2\over 6}T_n^2+{e^2\over 4}(V_n-V_m)^2\right].
\label{edensity2}
\end{equation}

On the other hand, in perfect metallic reservoirs (where $\sigma\to
\infty$)
the term ${\bf j}\!\cdot\!{\bf E}\sim {\bf j}^2/\sigma$ can be neglected in
Eq.\ (\ref{econserv}).
Integration of this equation over the volume of the $n$th metallic
reservoir gives the total amount of  energy transferred to (or absorbed from)
the
lattice in this reservoir, $Q_n=\int d{\bf r}q({\bf r})=-\int_{L_n}
d{\bf s}\!\cdot\!\hat {D}\nabla\Upsilon$.
In the particular case of thermal equilibrium between the reservoirs, i.e.,
$T_n=T$, $n=1,\ldots , N$,
we can use Eq.\ (\ref{edensity2}) to get the Joule heat
in the $n$th reservoir,
\begin{equation}
Q_n={1\over 2}\sum_mG_{nm}(V_n-V_m)^2.
\label{Jheat}
\end{equation}
For a two-terminal conductor, $(V_1-V_2)^2=V^2$,
$G_{12}=G_{21}=G$, we have $Q_1=Q_2=GV^2/2$, while the total Joule
heat is $Q_1+Q_2=IV$. We see in this case that the  heat contributions
released on each side of the two-terminal conductor are
equal.\cite{Levinson}
This general conclusion holds for
an arbitrary shape of the conductor and arbitrary disorder distribution.
This fact is a consequence of 
electron-hole symmetry.

The following simple analysis of the Eq.\ (\ref{Jheat})
exhibits its physical meaning. On one hand,
the total amount of Joule heat, 
${1\over 2}\sum_{nm}G_{nm}(V_n-V_m)^2=-\sum_nI_nV_n
=\int d{\bf r}{\bf j}\!\cdot\!{\bf E}$, is
simply equal to the total work done by the electric field on the system.
On the other hand,
the value $\frac{1}{2}e^2\nu_{F}(V_n-V_m)^2$
can be thought of as the gauge invariant difference of energy
densities $\Lambda$ (i.\ e.\ minus the density of the potential energy, 
$e^2\nu_F(V_m-V_n)V_n$) applied to the contacts of the conductor. 
Then the energy transport coefficients, $G_{nm}/e^2\nu_F$,
are determined by the conductance matrix. The last fact
is a manifestation of the Wiedemann-Franz law, which holds for diffusive
conductors (together with Eqs.\ (\ref{edensity2}) and (\ref{Jheat}))
in the cases of cold and hot electrons, as soon as the
electron-phonon interaction is weak enough. To show
the Wiedemann-Franz law in its usual form, we consider the thermal
transport in multiterminal conductors in the absence of charge
transport, $V_n=0,$ $n=1,\ldots ,N$. In this case we can use again
the Eq.\ (\ref{edensity2}) to calculate the thermal current $Q_n$,
\begin{equation}
Q_n={{\pi^2}\over {6e^2}}\sum_mG_{nm}T_m^2.
\label{Q1}
\end{equation}
In particular, close to thermal equilibrium, $T_m=T+\Delta T_m$,
we have
\begin{equation}
Q_n={{\pi^2T}\over {3e^2}}\sum_mG_{nm}\Delta T_m,\quad\Delta T_m
\ll T,
\label{Q2}
\end{equation}
where ${{\pi^2T}\over {3e^2}}G_{nm}$ is the thermal conductance matrix.
This is now the Wiedemann-Franz law in its usual form.

\subsection{Multiterminal spectral density of noise}
\label{density}

In this section we derive the general formula
for the multiterminal spectral density of shot noise
in the case of arbitrary electron-phonon interaction.
We multiply the first of Eqs.\ (\ref{diffb}) by
$\nabla\phi_n$
and integrate it over the volume of the conductor. Then we evaluate
the first term in {\em lhs} of the equation integrating by parts and
using the second of Eqs.\ (\ref{diffb}),
$\int d{\bf r}\nabla\phi_n\!\cdot\!\delta {\bf j}=\oint d{\bf s}\!
\cdot\!\delta {\bf j}\phi_n$. Taking into account the boundary
conditions (\ref{boundb}) for $\delta {\bf j}$ and (\ref{character2}) for $
\phi_n$ we get
$\int d{\bf r}\nabla\phi_n\!\cdot\!\delta {\bf j}=\delta I_n$.
Integration by parts in the second term of the {\em lhs} of this
equation gives $\int d{\bf r}\nabla\phi_n\!\cdot\!\hat\sigma\nabla\delta
V=\oint d{\bf s}\!\cdot\!\hat\sigma\nabla\phi_n\delta V=-\sum_kG_{nk}\delta
V_k(t)$,
where we used  Eqs.\ (\ref{character}, \ref{character2})
for $\phi_n$, the boundary condition (\ref{boundb})
for $\delta V$, and 
(\ref{conductance}) for the conductance matrix $G_{nm}$.
This leads us to the solution of the Langevin equation
(\ref{diffb}) in terms of characteristic potentials:
\begin{equation}
\delta\tilde {I}_n\equiv\delta I_n-\sum_mG_{nm}\delta V_m=
\int d{\bf r}\nabla\phi_n\!\cdot\!\delta {\bf j}^s.
\label{aux}
\end{equation}
Now, using the correlator (\ref{corr2}) for the Langevin sources $\delta
{\bf j}^s$,
we express the generalized multiterminal spectral density $S_{nm}$
defined as
\begin{equation}
S_{nm}=\int\limits_{-\infty}^{\infty}dt\langle\delta\tilde {I}_n(t)\delta
\tilde {I}_m(0)\rangle
\label{spectr2}
\end{equation}
in terms of characteristic potentials,
\begin{equation}
S_{nm}=\int d{\bf r}\nabla\phi_n\!\cdot\!\hat\sigma\nabla\phi_m\Pi ,
\label{MSD}
\end{equation}
with the properties: $S_{nm}=S_{mn}$, $\sum_nS_{nm}=0$, and $S_{nn}>0$.
In  equilibrium $\Pi({\bf r})=2T$, and (\ref{MSD}) together with  Eq.\
(\ref{conductance2}) lead to the result  for the thermal noise,
\begin{equation}
S_{nm}=-2G_{nm}T,
\label{thermal}
\end{equation}
which is again a manifestation of the fluctuation-dissipation theorem.

The formula (\ref{MSD}) is one of the central results of the paper.
It is valid for elastic and inelastic scatterings
and for an arbitrary multiterminal diffusive conductor.
The relation of $S_{nm}$ to the measured noise
is now as follows.
If, say,  the voltages are fixed, then $\delta I_n(t)=\delta
\tilde {I}_n(t)$, and
the matrix $S_{nm}=S_{nm}^c$ is directly measured. On the other
hand, when currents are fixed, $S_{nm}$ can be obtained
from the measured voltage correlator
$S^v_{nm}=\int_{-\infty}^{\infty}dt\langle
\delta V_n(t)\delta V_m(0)\rangle$ by tracing it with conductance matrices:
$S_{nm}=\sum_{n^{\prime}m^{\prime}}G_{nn^{\prime}}G_{mm^{\prime}}S_{
n^{\prime}m^{\prime}}^v$. The physical interpretation of
(\ref{MSD}) becomes now transparent:
$\Pi$ describes the broadening of the distribution
function (effective temperature) that is induced by the voltage applied
to the conductor and $\hat\sigma\Pi$
can thus be thought of as a local noise {\it source} (see the discussion
following the Eq.\ (\ref{corr2})), while $\phi_n$
can be thought of as the {\it probe} of this local noise.
In particular, this means that only $S_{nm}$ is of physical
relevance but not the current or voltage correlators themselves.

Let us consider now one important application of Eq.(\ref{MSD}).
In an experiment one can measure the local broadening $\Pi$ of the
nonequilibrium distribution function $f_0$ (effective noise temperature
$\Pi$)
at some point ${\bf r}=
{\bf r}_0$ on the
surface of the conductor by measuring the voltage fluctuations
in a noninvasive voltage probe. This is an open contact with a
small area on the surface of the
conductor around the point ${\bf r}={\bf r}_0$. The contact is not attached
to the reservoir
so that it does not cause the equilibration of the electron
gas, and as a result $\Pi =const$ around the point ${\bf r}_0$. Then,
(\ref{MSD}) can be rewritten as follows:
$S=\int d{\bf r}\nabla\phi\!\cdot\!\hat{\sigma}\nabla\phi\Pi =\Pi
({\bf r}_0)\int d{\bf r}\nabla\phi\!\cdot\!\hat{\sigma}\nabla\phi$, where $
\phi$ is the characteristic
potential corresponding to the noninvasive probe.
Using Eqs.\ (\ref{conductance2})
we get $S=R^{-1}\Pi ({\bf r}_0)$, where $R$ is the resistance of the contact
which comes from the volume around ${\bf r}_0$. Finally, taking into account
(\ref{aux}, \ref{spectr}) and the fact, that there is no current through
the voltage probe, $\delta I=0$, we obtain
\begin{equation}
S^v=R\Pi ({\bf r}_0).
\label{measure}
\end{equation}
This means that $\Pi$ can be directly measured which gives an important
information about nonequilibrium processes in the conductor.
Eq.\ (\ref{measure}) resembles the fluctuation-dissipation theorem.
This is so because there is no transport through the noninvasive
probe, and therefore one can think of the probe as being in local
equilibrium
with the effective temperature $\Pi$.
For this reason our consideration restricted to the diffusion regime
can in principle be  applied to the case of the tunnel coupling between
the probe and conductor. A possible experiment that could measure
shot noise at local tunneling contacts is discussed in detail in
Ref.\ \onlinecite{Gram}.
The above result can be easily generalized to take into account
the equilibration by the contact (see  Sec. \ref{thermtrans}).
There will be then an additional
noise suppression factor in  Eq.\ (\ref{measure}).

We note that (\ref{MSD}) together with
Eqs.\ (\ref{kineq2a}, \ref{kineq2b}, \ref{bound2}) for the average
distribution
function $f$ and Eqs.\ (\ref{character}, \ref{character2}) for the
characteristic potentials can serve as a starting point for
numerical evaluations of $S_{nm}$.
For purely elastic scattering as well as for hot electrons
it is even possible
to get closed analytical expressions for $S_{nm}$ as we will show next.
The physical conditions for different transport regimes are
discussed in Ref.\ \onlinecite{Nagaev2}.
In Sec.\ \ref{elastic} and \ref{hot} we will consider 
the charge transport ($T_n=T$, $n=1\ldots N$), and in the Sec.\
\ref{thermtrans} we will discuss the thermal transport
($V_n=0$, $n=1\ldots N$).

\section{Elastic scattering}
\label{elastic}

In the case of purely elastic scattering, $I[f]=0$,
the average distribution function satisfies the diffusion equation
\begin{equation}
\nabla\!\cdot\!\hat\sigma\nabla f_0=0,
\label{diff3}
\end{equation}
and the boundary conditions (\ref{bound2}) with $T_n =T$
(i.\ e.\ in the charge transport regime). Using this equation
one can prove (see Appendix \ref{A}) that for elastic scattering
cross correlations ($n\neq m$) are always negative,
in agreement with the general conclusion of
Ref.\ \onlinecite{Buttiker2}.

Eq.\ (\ref{diff3}) can be solved in terms of $\phi_n$:
$f_0=\sum_n\phi_nf_T(\varepsilon -eV_n)$.
Substituting this solution
into Eq.\ (\ref{corr2}) and using the sum rule (\ref{sum}) for $\phi_n$,
we can express  $\Pi$ in the following form,\cite{comparison}
\begin{equation}
\Pi =2\int d\varepsilon\sum_{k,l}\phi_k\phi_lf_T(\varepsilon -eV_
k)[1-f_T(\varepsilon -eV_l)].
\label{Pi}
\end{equation}
Performing the integration
over $\varepsilon$ we obtain,
\begin{equation}
\Pi =e\sum_{k,l}\phi_k\phi_l\left(V_k-V_l\right)\coth\left[{{e\left
(V_k-V_l\right)}\over {2T}}\right],
\label{temp}
\end{equation}
which in combination with Eq.\ (\ref{MSD}) gives the final expression
for $S_{nm}$ which is valid for purely elastic scattering. 
Eq.\ (\ref{temp}) describes
the crossover from the shot noise in multiterminal diffusive
conductors ($T\to 0$),
\begin{equation}
\Pi =e\sum_{k,l}\phi_k\phi_l|V_k-V_l|,
\label{temp2}
\end{equation}
to the equilibrium Johnson-Nyquist noise given by (\ref{thermal}).

\subsection{Universality of noise}
\label{universality1}

Now we are in the position to generalize the proof of universality
of the $1/3$-suppression of shot noise
\cite{Beenakker1,Nagaev1,Jong1,Nazarov} to the case of
an arbitrary {\it multiterminal\/} diffusive conductor.
To be specific we choose
$V_n=0$,
for $n\neq 1$, i.e. only contact $n=1$ has a non-vanishing voltage.
Then, using the sum rule (\ref{sum}) for $\phi_n$,
we get
\begin{eqnarray}
\Pi =2e\phi_1(1-\phi_1)V_1\coth\left(eV_1/2T\right) \nonumber \\
+2T(1-\phi_1)^2+2T\phi_1^2 \, .\nonumber
\end{eqnarray}
To get $S_{1n}$ we substitute this equation into
(\ref{MSD}) and evaluate the  first term as follows:
$\int d{\bf r}\nabla\phi_n\!\cdot\!\hat\sigma\nabla\phi_1\phi_1(1-\phi_1)=
\oint d{\bf s}\!\cdot\!\hat\sigma \nabla\phi_n(\phi_1^2/2-\phi_1^3/3)=
-G_{1n}/6$, where
we used (\ref{character}, \ref{character2}).
Similarly, for the integrals in the second and third term
we get: $\int d{\bf r}\nabla\phi_n\!\cdot\!\hat\sigma\nabla\phi_1(1-\phi_1
)^2=\int d{\bf r}\nabla\phi_n\!\cdot\!\hat\sigma\nabla\phi_1\phi_1^2=-G_{1
n}/3$.
Combining these results we arrive at
\begin{equation}
S_{1n}=-{1\over 3}G_{1n}\left[4T+eV_1\coth\left(eV_1/2T\right)\right].
\label{unicold}
\end{equation}
When $V_1=0$ we get $S_{1n}=-2G_{1n}T$, and the formula
for the Johnson-Nyquist noise is recovered. When $T=0$, we
express $S_{1n}$ in terms of outgoing currents, $I_n = G_{1n}V_1$:
\begin{eqnarray}
S_{1n} & = & -{1\over 3}e|I_n|,\quad n\neq 1,\nonumber \\
S_{11} & = & {1\over 3}e|I_1|.
\label{unicold2}
\end{eqnarray}
We note that above derivation is valid for arbitrary impurity distribution
and shape of the conductor, and for an arbitrary electron spectrum 
(band structure).  In this sense the suppression
factor ${1\over 3}$ is indeed universal. This generalizes the known
universality of a two--terminal conductor
\cite{Nazarov} to a multiterminal geometry.

Finally, we mention here some 
inequalities (derived in Appendix \ref{A}), which can be used to estimate
the spectral density $S_{nm}$ in the $T=0$ limit.
First, the correlations are bounded from below,
\begin{equation}
S_{nn}\geq{1\over 3}e|I_n|,
\label{below}
\end{equation}
but due to the nonlocality of the noise (see the discussion in
Sec.\ \ref{nonlocality}) there can be no upper bound
in terms of the current $I_n$ through the
same contact. In other words, the current $I_n$ flowing through
the $n$-th contact creates the noise $\frac{1}{3}eI_n$ in this contact.
However, other contacts also contribute to the noise in the $n$-th contact,
and 
this contribution is not universal and
makes the noise arbitrarily larger compared to the value $\frac{1}{3}eI_n$.
Nevertheless, we can write: $\Pi<\max\{|V_k-V_l|\}$, $k,l=1,\ldots ,N$,
which gives the rough
estimate
\begin{equation}
S_{nn} < e|G_{nn}|\max\{|V_k-V_l|\}.
\label{rough}
\end{equation}
In contrast, the cross correlations possess an upper bound,
\begin{equation}
|S_{nm}|\leq{1\over 2}(S_{nn}+S_{mm}).
\label{above}
\end{equation}
$S_{nm}$ vanishes when the $n$-th and $m$-th contacts are completely 
disconnected.

\subsection{Wide and star-shaped conductors}
\label{wide}

Next we specialize to two experimentally important cases.
First we consider a multiterminal conductor
of a star  geometry with $N$ long leads
(but with otherwise arbitrary shape)
which join each other at a small crossing
region (see Fig.\ \ref{noisefig}b).
The resistance of this region is assumed to be
much smaller
than the resistance of the leads.
In the second case the contacts are connected through a wide
region (see Fig.\ \ref{noisefig}c),
where again the
resistance of the conductor comes mainly from the regions near
the contacts, while the resistance of the wide region is
negligible.

Both shapes are characterized by the requirement
that $w/L\ll 1$, where $w$ and $L$ are the characteristic
sizes of the contact and of the entire conductor, resp.
In both cases the conductor can be divided (more or less arbitrary)
into $N$ subsections $\Gamma_k$, $k=1,\ldots ,N$,
associated
with a particular contact so that the potential $V$ is
approximately constant (for $w/L\ll 1$) on the dividing surfaces
$C_k$. Each subsection then can be thought as a two-terminal
conductor with the corresponding characteristic potential $\theta_
k({\bf r})$,
\begin{equation}
\nabla\!\cdot\!\hat\sigma\nabla\theta_k\! =\! 0,\quad
d{\bf s}\!\cdot\!\hat\sigma\nabla\theta_
k|_S\! =\! \theta_k|_{L_k}\! =\! 0,\;\;\theta_k|_{C_k}\! =\! 1.
\label{theta}
\end{equation}
We will show now that both, the multiterminal
conductance matrices $G_{nm}$  and the spectral densities $S_{nm}$, can be
expressed in terms of the conductances $G_k$ of these subsections,
\begin{equation}
G_k=-\int\limits_{L_k}d{\bf s}\!\cdot\!\hat\sigma\nabla\theta_k =
\int\limits_{C_k}d{\bf s}
\!\cdot\!\hat\sigma\nabla\theta_k .
\label{subsect}
\end{equation}

Since each potential $\phi_n$ is approximately constant in the central
region
of the multiterminal conductor,  we can write,
\begin{equation}
\phi_n({\bf r})|_{C_k}=\alpha_n=const.,\quad\sum_n\alpha_n=1,
\label{alpha}
\end{equation}
for an arbitrary $k=1,\ldots ,N$, where the second equation follows from
the sum rule for $\phi_n$.
Comparing Eqs.\ (\ref{character}, \ref{character2}, \ref{alpha})
with the definition of $\theta_k$ (\ref{theta}), we
immediately obtain
\begin{equation}
\phi_n({\bf r})|_{{\bf r}\in\Gamma_k}=\alpha_n\theta_k({\bf r})+
[1-\theta_k({\bf r})]\delta_{nk}.
\label{phi}
\end{equation}

The calculation of $G_{nm}$ and $S_{nm}$ is now straightforward.
We substitute (\ref{phi}) into (\ref{conductance}) and use 
Eq.\ (\ref{subsect}) to get
\begin{equation}
G_{nm}=(\alpha_m-\delta_{nm})G_n,\quad \alpha_m=G_m/G,
\label{conductance3}
\end{equation}
where $G\equiv\sum_nG_n$, and
the equation for $\alpha_m$ follows from $\sum_nG_{nm}=0$.
Substituting (\ref{phi}) into (\ref{MSD}) and
applying similar arguments as above in the proof
of the 1/3-suppression we find the explicit expressions
(for details of the derivation see  Appendix \ref{B})
\begin{eqnarray}
S_{nm}&=&{1\over 3}e\sum\limits_k\alpha_n\alpha_k(J_k+J_n)(\delta_{
nm}-\delta_{km})-{2\over 3}G_{nm}T,\nonumber\\J_n&=&\sum\limits_lG_
l(V_n-V_l)\coth\left[{{e(V_n-V_l)}\over {2T}}\right].
\label{MSD2}\end{eqnarray}
We note that this result is a consequence of above approximation (\ref{alpha}).
Comparing the resistance of the subsections to the resistance
of the central region of the conductor (which is neglected) we find
that the corrections to Eq.\ (\ref{alpha}) and consequently to 
Eq.\ (\ref{MSD2}) are of order $w/L$ in 3D and for a star geometry in 2D,
and up to corrections of order $[\ln (L/w)]^{-1}$
for wide conductors in 2D.

In principle, (\ref{MSD2}) and (\ref{conductance3}) allow us
to calculate the noise for arbitrary voltages and temperature, but
for illustrative purposes we consider
the simple case of a cross-shaped conductor
with four equivalent leads
(i.e. $\alpha_n=1/4$) and $T=0$.
Suppose the voltage is applied to only one contact, say
$V_1>0$,  $V_{n\neq 1}=0$, and $I=-I_1=3I_{n\neq 1}>0$.
Then, from (\ref{MSD2})
we obtain: $S_{11}={1\over 3}eI$, $S_{12}=S_{13}=S_{14}=-{1\over 9}
eI$, all being in agreement with
the universal $1/3$-suppression proven above. Then,
$S_{22}=S_{33}=S_{44}={2\over 9}eI$, and $S_{23}=S_{24}=S_{34}=-{1\over {
18}}eI$.
These numbers seem to be new \cite{comment1} and it would be interesting
to test them experimentally.

\subsection{Nonlocality and exchange effect}
\label{nonlocality}

We are now in the position to address the issue
of {\em non-locality} and {\em exchange} effect in shot
noise ($T=0$) in multiterminal conductors.
For this we consider for instance a star geometry
and assume that the
current enters the conductor through
the $n$-th contact, i.e. $I_n=-I$, and leaves it through
the $m$-th contact, i.e. $I_m=I$,
while the other contacts are  open, i.e. $I_k=0$ for $k\neq n,m$.
{}From (\ref{conductance3}) we obtain for the conductance
$G_nG_m/(G_n+G_m)$ (two contacts are in series), and we see that
it does not depend on the other leads, which simply reflects
the {\em local} nature of diffusive transport.
However, contrary to one's first expectation,
this locality does {\em not} carry over to the noise
in general. Indeed, from (\ref{MSD2})
it follows that
$S_{nm}=-{1\over 3}(\alpha_n+\alpha_m)eI$.
The additional
suppression factor $0<\alpha_n+\alpha_m<1$ for $N>2$ reflects the
{\em non-locality} of the current noise.
For instance, for a cross with $N=4$ equivalent leads we have
$\alpha_m=\alpha_n=1/4$, and thus $S_{nm}=-{1\over 6}eI$.
An analogous reduction factor was obtained in
Ref.\ \onlinecite{Beenakker1} under a different point of view.
Hence, one cannot disregard
open contacts simply because no current is flowing through them;
on the contrary, these open contacts which are still connected to
the reservoir induce  equilibration of the electron gas
and thereby reduce its current noise.
We emphasize that this non-locality is a classical
effect in the sense that no quantum phase interference is involved
(phase coherent effects are {\it not\/} contained
in our Boltzmann approach). On the other hand, 
the origin of this non-locality
can be traced back to the non-linear dependence
of $\Pi$ on the distribution $f$ in (\ref{corr2}),
which is a consequence of the Pauli exclusion principle.

Next we discuss exchange effects \cite{Buttiker2}
in a four terminal conductor.
According to Blanter and B\"uttiker \cite{Blanter} they can be
probed by measuring $S_{13}$  in three ways:
$V_n=V_0\delta_{n2}$ (A), $V_n=V_0\delta_{n4}$ (B), and
$V_n=V_0\delta_{n2}+V_0\delta_{n4}$ (C). Then we take
$\Delta S_{13}=S_{13}^C-S_{13}^A-S_{13}^B$
as a measure of the exchange effect.
This experiment is analogous to the experiment of
Hanbury Brown and Twiss in optics. \cite{Brown}
It measures the interference of electrons coming
from mutually incoherent sources, which is caused by the
indistinguishability of the electrons. 
Naively, one might expect that this interference
effect averages to zero in diffusive conductors.
However, it comes now as some surprise that in our semiclassical
Boltzmann approach
$\Delta S_{13}$ turns out to be non-zero in general
and can even be of the order of the shot noise itself.
Again, the reason for that is that $\Pi$ is non-linear in $f_0$
(see (\ref{corr2})), which is the consequence of the Pauli exclusion
principle. So, the value $\Pi^C-\Pi^A-\Pi^B$
which enters $\Delta S_{13}$ is not necessarily zero.
Indeed, while exchange effects vanish for cross
shaped conductors (in agreement with Ref.\ \onlinecite{Blanter}
up to corrections
of order $w/L$ which are neglected in our approximation),
it is not so for an H-shaped conductor (see Fig.\ \ref{noisefig}d).
Calculations similar to those leading to
(\ref{MSD2}) give for this case:
\begin{equation}
\Delta S_{13}={1\over {24}}{{eV_0G^2G_0}\over {(G+4G_0)^2}},
\label{exchange}
\end{equation}
where  $G_n=G/4$ are the  conductances (all being equal) of the
outer four leads, while the conductance of the  connecting wire
in the middle
is denoted by $G_0$.
This exchange term $\Delta S_{13}$ vanishes
for $G_0\to\infty$, because then the case of a simple cross is
recovered, and also
for $G_0\to 0$, because then the $1$-st and $3$-rd
contacts are  disconnected.
$\Delta S_{13}$ takes on its maximum value
for $G_0=G_n$ and becomes equal to ${1\over {60}}eI^A$,
where $I^A$ is the current
through the $2$-nd contact for case (A).

Although $\Delta S_{13}$ is positive in the example considered above 
this is not the case in general.
For an arbitrary four--terminal geometry of the conductor
the exchange effect can be expressed in terms of characteristic
potentials:
\begin{equation}
\Delta S_{nm}=-4eV_0\int d{\bf r}\nabla\phi_n\!\cdot\!\hat\sigma\nabla\phi_m
\phi_k\phi_l,
\label{exchange2}
\end{equation}
where all indices
are different. From this general formula it follows that
$\Delta S_{nm}=\Delta S_{kl}$, and $\Delta S_{nm}+\Delta S_{nl}+\Delta
S_{nk}=0$. The last equation means
that the exchange effect can change sign, i.e.\ cross correlations can be
either suppressed or enhanced. 

On the other
hand, the set-up can be slightly modified: instead of cross
correlations,
the noise density in one of the contacts of a multiterminal ($N>2$) diffusive
conductor is measured, say $S_{11}$, while the
electrons are injected through the contacts 2 (A), 3 (B),
and 2 and 3 (C). Again,
$\Delta S_{11}=S_{11}^C-S_{11}^A-S_{11}^B$ is a measure of the exchange
effect.
Then, it follows from (\ref{exchange2}) that
\begin{equation}
\Delta S_{11}=-4eV_0\int d{\bf r}\nabla\phi_1\!\cdot\!\hat\sigma\nabla\phi_1
\phi_2\phi_3<0,
\label{exchange3}
\end{equation}
i.e.\ the correlations are always suppressed due to the
interference effect, which is a direct manifestation of the Pauli principle.
In the particular case of star-shaped conductors we have
\begin{equation}
\Delta S_{11}=-\frac{4}{3}eV_0\frac{G_1G_2G_3}{G^2},\quad
G=\sum_{n=1}^{N}G_n.
\label{exchange4}
\end{equation}
The suppression of noise due to the interference of mutually
incoherent electrons was recently observed in an experiment with a
ballistic electron beam splitter. \cite{Tarucha} 
We have shown here that  this
effect is also observable in mesoscopic diffusive conductors.

\section{Hot electrons}
\label{hot}

We consider now the case of ``hot" electrons
where $I_{ee}\neq 0$, but still $I_{e-ph}=0$, and we assume that
electron-electron scattering  is sufficiently strong
to cause thermal equilibration of
the electron gas
(i.e. $l_{ee}=\sqrt {D\tau_{ee}}\ll L$,
where $D$ is the diffusion coefficient and $\tau_{ee}$
the electron-electron relaxation time).
The average distribution then assumes the Fermi-Dirac
form:
\begin{equation}
f_0(\varepsilon ,
{\bf r})=\left\{1+\exp\left[{{\varepsilon -eV({\bf r})}\over {T_e({\bf r})}}
\right]\right\}^{-1},
\label{distribution}
\end{equation}
with the local electron temperature $T_e({\bf r})$.
Substituting this $f_0$ into (\ref{corr2}) we immediately get
$\Pi ({\bf r})=2T_e({\bf r})$. On the other hand, from Eq.\ (\ref{edensity})
it follows that $(T_e({\bf r}))^2=(6/\pi^2\nu_F)\Upsilon ({\bf r})$,
where $\Upsilon ({\bf r})$ is given by (\ref{edensity2}) with $T_n =T$
(i.\ e.\ in the charge transport regime).
Thus, we finally obtain
\begin{equation}
\Pi =2T\left[1+2\sum_{n,m}\phi_n\phi_m(\beta_n\! -\!\beta_m)^2\right]^{{
1\over 2}},\;
\beta_n\! =\! {{\sqrt 3eV_n}\over {2\pi T}},
\label{temp3}
\end{equation}
which in combination with (\ref{MSD}) gives the general solution
for the case of hot electrons.
We would like to note here that
the cross correlations are always negative
also in the case of hot electrons (the proof is given in the Appendix
\ref{A}).

\subsection{Universality of noise}
\label{universality2}

Next we show that the shot-noise suppression factor
$\sqrt 3/4$ (Refs.\ \onlinecite{Nagaev2,Kozub})
for  hot electrons in a multiterminal conductor
is also {\it universal\/}.
As before
we can consider
the case  where the voltage is applied to only one contact:
$V_n=V_1\delta_{n1}$. Then
\begin{equation}
\Pi =2T\sqrt {1+4\beta^2\left(\phi_1-\phi_1^2\right)},
\label{temp4}
\end{equation}
where $\beta\equiv\beta_1$.
Using the relation
\begin{eqnarray}
2\sqrt {1-\Phi^2}\nabla\Phi =\nabla\left\{\arcsin \Phi +\Phi\sqrt {
1-\Phi^2}\right\},\nonumber \\
\Phi =\beta\left(1+\beta^2\right)^{-1/2}(2\phi_1-1),
\label{relation}
\end{eqnarray}
we transform the volume integral in
(\ref{MSD}) into a surface integral
and obtain the spectral density of noise:
\begin{equation}
S_{1n}=\! -G_{1n}T\left[1+\left(\beta\! +\!\frac{1}{\beta}\right)\arctan
\beta \right],\;
\beta ={{\sqrt 3eV_1}\over {2\pi T}}.
\label{unihot}
\end{equation}
This expression describes the crossover from
the thermal noise
($\beta\ll 1$) given by (\ref{thermal})
to the transport noise $(\beta\gg 1$)
\begin{eqnarray}
S_{1n} & = & -{{\sqrt 3}\over 4}e|I_n|,\quad n\neq 1,\nonumber \\
S_{11} & = & {{\sqrt 3}\over 4}e|I_1|.
\label{unihot2}
\end{eqnarray}
This general result shows that
in the case of hot electrons
the shot-noise suppression factor $\sqrt 3/4$ is indeed universal,
i.e. it does not depend on the shape of the multiterminal diffusive
conductor nor on its disorder distribution.\cite{comment2}

The origin of this universality becomes clear from the following
argumentation.
We have seen that the distribution of the
effective noise temperature $\Pi ({\bf r})$ for the case of hot electrons is
controlled by the transport equations for the energy, (\ref{econserv}, \ref{bound3})
through the heat density $\Upsilon ({\bf r})$.
The spectral density of noise, in turn,
is given by $\Pi ({\bf r})$ through the  transport equations for charge,
(\ref{diffb}, \ref{boundb}). On the other hand, according to
the Wiedemann-Franz law both, the energy and charge transports, are
determined by the same kinetic coefficients, namely, by $\hat\sigma ({\bf
r})$.
Thus, the physical origin of the universality of the
suppression factor $\sqrt 3/4$ can be traced back to the
Wiedemann-Franz law.
Conversely, a violation of the Wiedemann-Franz law will
cause deviations from universality.

We would like to note here that
the universality of the noise (for cold and hot electrons)
has been proven here for the case where the voltage is applied to
only one contact of a multiterminal diffusive conductor which
made it possible to express the spectral densities $S_{nm}$
in terms of conductances $G_{nm}$. This is no longer possible in general.
Nevertheless, in the case of a 2D geometry and isotropic
conductivity,
$\sigma_{\alpha\beta}({\bf r})=\sigma ({\bf r})\delta_{\alpha\beta}$,
both $G_{nm}$ and $S_{nm}$
are of the same  universality class. Indeed, one can easily see that
they are invariant
under conformal transformation of coordinates. 
$G_{nm}$ and $S_{nm}$ can be expressed in terms of  characteristic
potentials $\phi_n$, which satisfy the conformal invariant
diffusion equation (\ref{character}) and boundary conditions
(\ref{character2}).
Moreover, the combination $d{\bf r}\nabla\phi_n\!\cdot\!\nabla\phi_
m$ does not change with the
conformal transformation of coordinates, which  finally makes
the integrals for $G_{nm}$ (\ref{conductance2}) and for $S_{nm}$
(\ref{MSD}) conformal invariant.

We close this section by another
illustrative example. Let us consider again
a cross-shaped conductor with four equivalent leads,
\cite{comment3}
$G_n=G/4$, at $T=0$ and where we choose
$V_n=V_1\!\delta_{1n}$, $I=-I_1=3I_{n\neq 1}>0$. We then find
$S_{11}={{\sqrt 3}\over 4}eI$,
and $S_{1n}=-{1\over {4\sqrt 3}}eI$,
for $n\neq 1$,
while
$S_{nn}=\left({{35\sqrt 3}\over {108}}-{2\over {3\pi}}\right)eI$,
and
$S_{n\neq m}=-\left({{13\sqrt 3}\over {108}}-{1\over {3\pi}}\right
)eI$,
for $n,m\neq 1$.
These new numbers are consistent with the universal factor $\sqrt
3/4$.

\section{Noise induced by thermal transport}
\label{thermtrans}

In this section we address a new phenomenon, namely the current
noise in multiterminal diffusive conductors in the presence of
thermal transport. We assume no energy relaxation
in the conductor due to phonons, $I_{e-ph}=0$, and no voltage is applied
to the contacts, $V_n=0$, $n=1,\ldots ,N$, which are kept
in  equilibrium at different temperatures $T_n$. The thermal transport
is considered in  Sec. \ref{energy}, where the outgoing thermal
currents $Q_n$ are calculated (see Eqs.\ (\ref{Q1}) and (\ref{Q2})).
We turn now to the calculation of the spectral density of noise.

\subsection{Elastic scattering}
\label{coldheat}

To calculate $\Pi$ we need to know the distribution function
$f_0$. It obeys the Eq.\ (\ref{diff3}) with the boundary conditions
(\ref{bound2}) in the contacts. The solution then
reads explicitly,
\begin{equation}
f_0=\sum_n\phi_nf_{T_n},
\label{distribution2}
\end{equation}
and with the help of (\ref{sum}) we get,
\begin{eqnarray}
\lefteqn{\Pi =\sum_{kl}\phi_k\phi_lZ_{kl},}\nonumber \\
& & Z_{kl}=T_kT_l\int\limits^{\infty}_{
-\infty}ds\left[1-\tanh (T_ks)\tanh (T_ls)\right].
\label{Pi2}
\end{eqnarray}
This together with  Eq.\ (\ref{MSD}) gives the spectral density
of noise $S_{nm}$.

In  equilibrium $T_n=T$,  $Z_{kl}=2T$ and (\ref{Pi2})
and we find for the equilibrium noise, $S_{nm}=-2G_{nm}T$.
On the other hand, if for example $T_k\gg T_l$, then $Z_{kl}=(2\ln
2)T_k$.
We consider then two situations, where e.g. either
$T_1=T$ and $T_{n\neq 1}=0$, or $T_1=0$ and $T_{n\neq 1}=T$.
In other words, only one contact is either heated
up to high enough temperature $T$ or cooled down to zero
temperature and the other cantacts are kept at the same temperature.
Using the sum rule for $\phi_n$
and carrying out the integration in (\ref{MSD}) we obtain for both
cases,
\begin{equation}
S_{1n}=-{2\over 3}(1+\ln 2)G_{1n}T.
\label{MSD3a}
\end{equation}
$S_{1n}$ for this two situations can be expressed in terms of
thermal currents $Q_n$
\begin{equation}
S_{1n}=\pm 4(1+\ln 2)(e/\pi )^2T^{-1}Q_n,
\label{MSD3b}
\end{equation}
with the sign depending on the
sign of $Q_n$.

\subsection{Hot electrons}
\label{hotheat}

We consider now the case of hot electrons, where 
$\Pi =2T_e$, while $T_e^2=\sum_k\phi_kT_k^2$
(see Eq.\ (\ref{edensity2})). Substituting this into
(\ref{MSD}) we get,
\begin{equation}
S_{nm}=2\int d{\bf r}\nabla\phi_n\!\cdot\!\hat{\sigma}\nabla\phi_m
\sqrt{\sum_k\phi_kT_k^2}.
\label{MSD4a}
\end{equation}
In particular, if the electron gas in the conductor is pushed
out of  equilibrium
by heating (or cooling)
one of the contacts (with $n=1$) while the other contacts are kept at the
temperature $T_n=T_2$, $n\neq 1$, the integral can be calculated explicitly,
and we have
\begin{equation}
S_{1n}=-{4\over 3}G_{1n}{{T_1^2+T_2^2+T_1T_2}\over {T_1+T_2}}.
\label{MSD4b}
\end{equation}
In the cases where $T_1=T\gg T_2$,
and $T_2=T\gg T_1$,  we obtain with the help of Eq.\ (\ref{Q2}) that
the spectral density $S_{1n}$ can be expressed in terms of thermal
currents:
\begin{equation}
S_{1n}=\pm 8(e/\pi )^2T^{-1}Q_n.
\label{MSD4c}
\end{equation}
Expressions (\ref{MSD3b}, \ref{MSD4c}) are analogous to 
Eqs.\ (\ref{unicold2}, \ref{unihot2})
and reflect the universality of the noise in the presence of thermal
transport.

\section{Conclusion}
\label{conclusion}

In conclusion, we have systematically studied the transport and noise in
multiterminal diffusive conductors. Applying a diffusion approximation to the
Boltzmann-Langevin kinetic equation we have derived the diffusion equations for
the distribution function and its fluctuations. We then solved these equations in
general terms of well defined ``characteristic potentials'' and we derived exact
formulas for the conductance matrix, energy transport coefficients and the
multiterminal spectral density of noise. In this way we have obtained the
following results. In both regimes of cold and hot electrons the shot noise turns
out to be universal in the sense	that it depends neither on the geometry of a
multiterminal conductor and the spectrum of carriers, nor on the disorder
distribution. We have studied the noise in the presence of thermal transport and
find that being expressed in terms of thermal currents it is also universal. We
believe that the origin of this universality lies in the fact that in the
diffusive regime the correlator of the local current densities (Langevin sources)
takes an equilibrium-like form of the fluctuation-dissipation theorem involving
an effective noise temperature. Thus, the transport and noise properties are
determined by the same conductivity tensor. One can surmise then that the proven
universality holds as long as the energy transport is governed by the
Wiedemann-Franz law.

The exchange effect is proven to be non-zero even within our semiclassical
Boltzmann approach. The exchange effect can change sign when measured in
cross-correlations, and (in agreement with the Pauli principle) it gives always
negative contribution to the auto-correlations. The exchange effect comes from a
non-linear dependence on the local distribution function. Similarly we show that
the same non-linearities are responsible for non-local effects such as the
suppression of shot noise by open leads even at zero electron temperature.

Finally, we have proposed a possible experiment which would allow one to locally
measure the effective noise temperature, and we have given new suppression
factors for shot noise in various geometries which can be tested experimentally.

\acknowledgments
We would like to thank M.\ B\"uttiker and Ch.\ Sch\"onenberger
for helpful discussions.
This work is supported by
the Swiss National Science
Foundation.

\appendix

\section{}
\label{A}

In this Appendix we derive some properties of the characteristic 
potentials $\phi_
n$, multiterminal conductance matrix $G_{nm}$ and spectral
density of shot noise $S_{nm}$.

First, we show that $\phi_n\geq 0$ in the conductor. According to
the boundary condition (\ref{character2}), being negative
$\phi_n$ would take on its minimum value at some point ${\bf r}={\bf r}_
0$
in the conductor.
If this happened  inside the conductor, then $\nabla\phi_n({\bf r}_0)=
0$, and
$\nabla\!\cdot\!\hat{\sigma }({\bf r}_0)\nabla\phi_n({\bf r}_0)>0$, because
$
\hat{\sigma}$ is a positive definite matrix.
This, however, would then
contradict  Eq.\ (\ref{character}). If $\phi_n$ took on its minimum value
on the open surface of the conductor (${\bf r}_0\in S$), then $\nabla_{
\parallel}\phi_n({\bf r}_0)=0$,
and according to Eq.\ (\ref{character2}) $\nabla_{\perp}\phi_n({\bf r}_
0)=0$. Again,
$\nabla\!\cdot\!\hat{\sigma }({\bf r}_0)\nabla\phi_n({\bf r}_0)>0$ in
contradiction with
Eq.\ (\ref{character}). Thus, we see
that the characteristic potentials cannot be negative, $\phi_n\geq
0$.

The sum rule (\ref{sum}) for $\phi_n$ follows from the
observation that the function $\phi ({\bf r})\equiv 1$ is a unique solution
of the diffusion equation $\nabla\!\cdot\!\hat{\sigma}\nabla\phi =
0$ with the boundary conditions
$d{\bf s}\!\cdot\!\hat{\sigma}\nabla\phi |_S=0$, and $\phi |_{L_m}
=1$, $\forall m$.  It follows from Eqs.\ (\ref{character}, \ref{character2})
that the function $\sum_n\phi_n({\bf r})$ obeys the same
equation and boundary conditions, and therefore
$\phi =\sum_n\phi_n({\bf r})=1$. 
This sum rule can also be deduced from the fact that
physical observables are invariant under a global shift
of the energy scale by a constant value.\cite{Buttiker3}

Now we prove that $G_{nm}>0$ for $n\neq m$. We note that the
multiplication of the integrand in (\ref{conductance}) by $\phi_m^2$ and
extension of the integral to the whole surface does not change the
integral, so that $G_{nm}=-\oint d{\bf s}\!\cdot\!\hat{\sigma}\nabla
\phi_n\phi_m^2$. Then, using
the Eq.\ (\ref{character}) we replace the surface integral by the
integral over the volume of the conductor, $G_{nm}=-2\int d{\bf r}
\nabla\phi_n\!\cdot\!\hat{\sigma}\nabla\phi_m\phi_m$.
Finally, we calculate this integral by parts and take into account
the boundary conditions (\ref{character2}) for $\phi_n$ to get 
for the conductance matrix,
\begin{equation}
G_{nm}=2\int d{\bf r}\nabla\phi_m\!\cdot\!\hat{\sigma}\nabla\phi_
m\phi_n,\quad n\neq m,
\label{conductance3a}
\end{equation}
from which it follows that $G_{nm}>0$. Interestingly, a similar procedure
for $n=m$ gives,
\begin{equation}
G_{nn}=-2\int d{\bf r}\nabla\phi_n\!\cdot\!\hat{\sigma}\nabla\phi_
n\phi_n,
\label{conductance3b}
\end{equation}
and $G_{nn}<0$.

Now we prove that in the case of elastic scattering
cross correlations $S_{nm}$, $n\neq m$, are always negative.
We calculate $S_{nm}$
in (\ref{MSD}) with $\Pi$ from (\ref{corr2}) integrating by parts,
\begin{eqnarray}
S_{nm}={1\over 2}\oint d{\bf s}\!\cdot\!\hat{\sigma }(\nabla\phi_
n\phi_m+\nabla\phi_m\phi_n)\Pi \nonumber \\
\mbox{}-{1\over 2}\int d{\bf r}\nabla\Pi\!
\cdot\!\hat{\sigma }(\nabla\phi_n\phi_m+\nabla\phi_m\phi_n).
\nonumber
\end{eqnarray}
{}From (\ref{bound2}) it follows that only contact surfaces
contribute to the first integral, and with
(\ref{conductance}) we get for the first term: $-G_{nm}(T_n+T_m)$. In the second
term we again calculate the integral by parts and use (\ref{corr2}) to get:
$-{1\over 2}\oint d{\bf s}\!\cdot\!\hat{\sigma}\nabla\Pi\phi_n\phi_
m-2\int\!\!\int d{\bf r}d\varepsilon\phi_n\phi_m\nabla f_0\!\cdot\!
\hat{\sigma}\nabla f_0$. According to (\ref{bound2})
and (\ref{character2}) the surface integral disappears and we arrive at
the following result
\begin{equation}
S_{nm}=-G_{nm}(T_n+T_m)-2\int\!\!\int d{\bf r}d\varepsilon\phi_n\phi_m\nabla
f_0\!\cdot\!\hat{\sigma}\nabla f_0,
\label{negative}
\end{equation}
for $n\neq m$, from which it follows that the cross correlations are
negative.

We apply similar arguments to prove that the cross correlations
are always negative also in the case of hot electrons.
We calculate the integral in $S_{nm}=2\int d{\bf r}\nabla\phi_n\!\cdot\!
\hat{\sigma}\nabla\phi_mT_e$ and use
the boundary conditions for $\phi_n$ and $T_e$ to get
$$
S_{nm}=-G_{nm}(T_n+T_m)+\int d{\bf r}\phi_n\phi_m\nabla\!\cdot\!\hat{\sigma}
\nabla T_e.
$$
Then we use $(T_e({\bf r}))^2=(6/\pi^2\nu_F)\Upsilon ({\bf r})$
and Eq.\ (\ref{econserv}) ($q=0$) to write
$T_e\nabla\!\cdot\!\hat{\sigma}\nabla T_e=-\nabla T_e\!\cdot\!\hat{
\sigma}\nabla T_e-3(e/\pi )^2\nabla V\!\cdot\!\hat{\sigma}\nabla V$.
Substituting this into
above equation we obtain
\begin{eqnarray}
& S_{nm} & =-G_{nm}(T_n+T_m) \nonumber \\
& - & \int d{\bf r}\phi_n\phi_m T^{-1}_e\left[\nabla T_
e\!\cdot\!\hat\sigma\nabla T_e+3(e/\pi )^2\nabla V\!\cdot\!\hat\sigma
\nabla V\right],
\label{negative2}
\end{eqnarray}
which shows the negativity of cross correlations.

The inequality (\ref{below}) can be derived as follows. First we note
that according to  Eq.\ (\ref{temp}) correlations grow with 
temperature. Therefore, without loss of generality we can put $T=0$
in the following. Then, one can easily see that
\begin{eqnarray}
\Pi =e\sum_{k,l}\phi_k\phi_l|V_k-V_l|\geq 2e\phi_n\sum_l\phi_l|V_
l-V_n| \nonumber \\
\geq 2e\phi_n\bigg|\sum_l\phi_l(V_l-V_n)\bigg|=2e\phi_n|V-V_n|.
\nonumber
\end{eqnarray}
We substitute this into (\ref{MSD}) and write another set of inequalities,
\begin{eqnarray}
S_{nn} & \geq &  2e\int d{\bf r}\nabla\phi_n\!\cdot\!\hat{\sigma}\nabla
\phi_n\phi_n|V-V_n| \nonumber \\
& \geq & 2e\bigg|\int d{\bf r}\nabla\phi_n\!\cdot\!\hat{
\sigma}\nabla\phi_n\phi_n(V-V_n)\bigg|.
\nonumber
\end{eqnarray}
Integrating by parts,
\begin{eqnarray}
\lefteqn{
2\int d{\bf r}\nabla\phi_n\!\cdot\!\hat{\sigma}\nabla\phi_n\phi_
n(V-V_n)
}
\nonumber \\
&& =\oint d{\bf s}\!\cdot\!\hat{\sigma}\nabla\phi_n\phi_n^2(V
-V_n)
-\int d{\bf r}\nabla V\!\cdot\!\hat{\sigma}\nabla\phi_n\phi_n^2
\nonumber \\
&&=-{1\over 3}\oint d{\bf s}\!\cdot\!\hat{\sigma}\nabla V\phi_n^3={
1\over 3}I_n,
\nonumber
\end{eqnarray}
we obtain expression (\ref{below}). The inequality Eq. (\ref{above})
immediately follows from
\begin{equation}
2|\nabla\phi_n\!\cdot\!\hat{\sigma}\nabla
\phi_m|\leq\nabla\phi_n\!\cdot\!\hat{\sigma}\nabla\phi_n+\nabla\phi_
m\!\cdot\!\hat{\sigma}\nabla\phi_m\, ,
\label{inequality}
\end{equation}
and evidently holds also for inelastic scattering.

\section{}
\label{B}

The derivation of (\ref{MSD2}) proceeds as follows.
We use (\ref{phi}) to replace the integral in (\ref{MSD}) over the volume
of the conductor by the sum of integrals over subsections $\Gamma_
k$,
\begin{equation}
S_{nm}=\!\sum_k(\alpha_n\! -\!\delta_{nk})(\alpha_m\! -\!\delta_{mk})\!\int
\limits_{\Gamma_
k}d{\bf r}\nabla\theta_k\!\cdot\!\hat{\sigma}\nabla\theta_k\Pi .
\label{MSD2a}
\end{equation}
Using the contraction
\begin{equation}
Z_{lq}=e(V_l-V_q)\coth\left[{{e(V_l-V_q)}\over {2T}}\right]
\label{MSD2b}
\end{equation}
we express $\Pi$ in terms of $\theta_k$,
\begin{eqnarray}
\Pi |_{\Gamma_k}=\sum_{lq}[\alpha_l\theta_k+(1-\theta_
k)\delta_{lk}]\nonumber \\
\mbox{}\times [\alpha_q\theta_k+(1-\theta_k)\delta_{qk}]Z_{lq}.
\label{MSD2c}
\end{eqnarray}
Then we substitute this expression into Eq.\ (\ref{MSD2a}), calculate
the integrals over $\Gamma_k$ with the help of Eqs.\ (\ref{subsect}),
\begin{eqnarray}
\int\limits_{\Gamma_k}d{\bf r}\nabla\theta_k\!\cdot\!\hat{\sigma}\nabla
\theta_k\theta_k^2=2\int\limits_{\Gamma_k}d{\bf r}\nabla\theta_k\!\cdot\!
\hat{\sigma}\nabla\theta_k\theta_k(1-\theta_k)\nonumber \\
\mbox{} =\int\limits_{\Gamma_k}d{\bf r}
\nabla\theta_k\!\cdot\!\hat{\sigma}\nabla\theta_k(1-\theta_k)^2={1\over
3}G_k,
\end{eqnarray}
and use the symmetry $Z_{lq}=Z_{ql}$ to get
\begin{eqnarray}
S_{nm} & = & {1\over 3}\sum_{klq}(\alpha_n-\delta_{nk})(\alpha_m-\delta_{
mk})\nonumber \\
\mbox{} & \times &
(\alpha_l\alpha_q+\alpha_l\delta_{qk}+\delta_{lk}\delta_{qk})G_
kZ_{lq}.
\label{MSD2d}
\end{eqnarray}
Finally, introducing the notation, $J_q=e^{-1}\sum_lG_lZ_{ql}$, and
using $Z_{kk}=2T$ and
$\alpha_k=G_k/G$, we carry out the summation over $k$ in (\ref{MSD2d})
to arrive at  (\ref{MSD2}) for $S_{nm}$.


\begin{references}
\bibitem[*]{Eugene} email: sukhorukov@ubaclu.unibas.ch
\bibitem[\dagger]{inst} On leave from
Institute of Microelectronics Technology,
Russian Academy of Sciences, Chernogolovka, 142432 Russia.
\bibitem[\ddagger]{Daniel} email: loss@ubaclu.unibas.ch
\bibitem{Schottky}
W.\ Schottky, Ann.\ Phys.\ (Leipzig) {\bf 57}, 541 (1918). 
\bibitem{Jongrev}
For a recent review, see
M.\ J.\ M.\ de Jong, and C.\ W.\ J.\ Beenakker, 
in {\em Mesoscopic electron
transport, NATO ASI, Series E: Applied Science},
eds. L.\ P.\ Kouwenhoven, G.\ Sch\"on, and L.\ L.\ Sohn
(KluwerDordrecht, in press).
\bibitem{Khlus} V.\ A.\ Khlus,
Zh.\ Eksp.\ Teor.\ Fiz.\ {\bf 93} 2179 (1987) 
[Sov.\ Phys.\ JETP {\bf 66}, 1243 (1987)].
\bibitem{Landauer}
R.\ Landauer, 
Physica D{\bf 38}, 226 (1989).
\bibitem{Lesovik}
G.\ B.\ Lesovik,
Pis'ma Zh.\ Eksp.\ Teor.\ Fiz.\ {\bf 49}, 513 (1989)
[JETP Lett.\ {\bf 49}, 592 (1989)].
\bibitem{Yurke}
B.\ Yurke, and G.\ P.\ Kochanski, 
Phys.\ Rev.\ B{\bf 41}, 8184 (1990).
\bibitem{Buttiker1}
M.\ B\"{u}ttiker, 
Phys.\ Rev.\ Lett.\ {\bf 65}, 2901 (1990).
\bibitem{Kulik}
I.\ O.\ Kulik, and A.\ N.\ Omel'yanchuk,
Fiz.\ Nizk.\ Temp.\ {\bf 10}, 305 (1984) 
[Sov.\ J.\ Low Temp.\ Phys.\ {\bf 10}, 158 (1984)].
\bibitem{Beenakker1}
C.\ W.\ J.\ Beenakker, and M.\ B\"{u}ttiker, 
Phys.\ Rev.\ B{\bf 46}, 1889 (1992).
\bibitem{Nagaev1}
K.\ E.\ Nagaev, 
Phys.\ Lett.\ A{169}, 103 (1992).
\bibitem{Jong1}
M.\ J.\ M.\ de Jong, and C.\ W.\ J.\ Beenakker, 
Phys.\ Rev.\ B{\bf 46}, 13400 (1992).
\bibitem{Nazarov}
Yu.\ V.\ Nazarov, 
Phys.\ Rev.\ Lett.\ {\bf 73}, 134 (1994).
\bibitem{Altshuler}
B.\ L.\ Altshuler, L.\ S.\ Levitov, and A.\ Yu.\ Yakovets,
Pis'ma Zh.\ Eksp.\ Teor.\ Fiz.\ {\bf 59}, 821 (1994)
[JETP Lett.\ {\bf 59}, 857 (1994)].
\bibitem{Jong}
M.\ J.\ M.\ de Jong, and C.\ W.\ J.\ Beenakker, 
Phys.\ Rev.\ B{\bf 51}, 16867 (1995).
\bibitem{Liefrink}
F.\ Liefrink, J.\ I.\ Dijkhuis, M.\ J.\ M.\ de Jong, 
L.\ W.\ Molenkamp, and H.\ van Houten,
Phys.\ Rev.\ B{\bf 49}, 14066 (1994).
\bibitem{Steinbach}
A.\ H.\ Steinbach, J.\ M.\ Martinis, and M.\ H.\ Devoret,
Phys.\ Rev.\ Lett.\ {\bf 76}, 3806 (1996).
\bibitem{Schoelkopf}
R.\ J.\ Schoelkopf, P.\ J.\ Burke, 
A.\ A.\ Kozhevnikov, D.\ E.\ Prober, and M.\ J.\ Rooks,  
Phys.\ Rev.\ Lett.\ {\bf 78}, 3370 (1997).
\bibitem{Schoenenberger}
M.\ Henny, H.\ Birk, R.\ Huber, C.\ Strunk, A.\ Bachtold, 
M.\ Kr\"{u}ger, and C.\ Sch\"{o}nenberger,  
Appl.\ Phys.\ Lett.\ {\bf 71}, 773 (1997).
\bibitem{Schoenenberger2}
M.\ Henny, S.\ Oberholzer, 
C.\ Strunk, and C.\ Sch\"{o}nenberger,  
cond-mat/9808042.
\bibitem{Shimitzu}
A.\ Shimitzu, and M.\ Ueda, 
Phys.\ Rev.\ Lett.\ {\bf 69}, 1403 (1992).
\bibitem{Nagaev2}
K.\ E.\ Nagaev, 
Phys.\ Rev.\ B{\bf 52}, 4740 (1995).
\bibitem{Kozub}
V.\ I.\ Kozub, and A.\ M.\ Rudin, 
Phys.\ Rev.\ B{\bf 52}, 7853 (1995).
\bibitem{Blanter}
Ya.\ M.\ Blanter, and M.\ B\"{u}ttiker, 
Phys.\ Rev.\ B{\bf 56}, 2127 (1997).
\bibitem{Kadomtsev}
B.\ B.\ Kadomtsev,
Zh.\ Eksp.\ Teor.\ Fiz.\ {\bf 32} 943 (1957)
[Sov.\ Phys.\ JETP {\bf 5}, 771 (1957)].
\bibitem{Kogan}
Sh.\ M.\ Kogan, and A.\ Ya.\ Shul'man,
Zh.\ Eksp.\ Teor.\ Fiz.\ {\bf 56} 862 (1969)
[Sov.\ Phys.\ JETP {\bf 29}, 467 (1969)].
\bibitem{Buttiker3}
M.\ B\"uttiker, 
J.\ Phys.: Condens.\ Matter {\bf 5}, 9361 (1993).
\bibitem{Buttiker2}
M.\ B\"{u}ttiker, 
Phys.\ Rev.\ B{\bf 46}, 12485 (1992).
\bibitem{Martin1}
Th.\ Martin, and R.\ Landauer, 
Phys.\ Rev.\ B{45}, 1742 (1992).
\bibitem{Jong2}
M.\ J.\ M.\ de Jong, and C.\ W.\ J.\ Beenakker, 
Physica A{\bf 230}, 219 (1996).
\bibitem{Landauer2}
For a critical discussion of the universality of noise,
see R.\ Landauer, Physica {\bf B}227, 156 (1996); preprint, (1998).
\bibitem{sukhor} 
E.\ V.\ Sukhorukov, and D.\ Loss,
Phys.\ Rev.\ Lett.\ {\bf 80}, 4959 (1998).
\bibitem{Gurevich}
V.\ L.\ Gurevich, and A.\ M.\ Rudin,
Phys.\ Rev.\ B{\bf 53}, 10078 (1996).
\bibitem{Naveh}
Y.\ Naveh, D.\ V.\ Averin, and K.\ K.\ Likharev,
Phys.\ Rev.\ Lett.\ {\bf 79}, 3482 (1997).
\bibitem{Nagaev3}
K.\ E.\ Nagaev, 
Phys.\ Rev.\ B{\bf 57}, 4628 (1998)
\bibitem{estimate}
This contribution is of order $(eV/\varepsilon_F)\cdot (l/L)$ 
in the diffusive regime, where the mean free path $l$
is much smaller than the size of the conductor $L$.
\bibitem{Landau}
E.\ M.\ Lifshitz, and L.\ P.\ Pitaevskii, 
{\em Physical Kinetics}
(Pergamon, Oxford, 1981), p.\ 330.
\bibitem{contactheat}
We mention here that this assumption is an 
idealization. In a real experiment the heating of the electron
gas in the reservoirs caused by the electron-electron interaction is 
hard to avoid, which makes the experimental observation
of the suppression factor of $1/3$ a difficult task.
On the other hand, the universality of noise 
(see Sec.\ \ref{universality1}) allows us to pull out the contact
surfaces $L_n$ far into the reservoirs. Thus a proper
choice of the metallic reservoirs
provides in principle the possibility to cool down the electron gas.
Therefore, we believe that the above assumption is justified.
The role of reservoirs is thoroughly studied in 
Ref.\ \onlinecite{Schoenenberger2}.
\bibitem{comment0}
A similar equation but in less general form has been used in Refs.\ 
\onlinecite{Nagaev2,Kozub}, and \onlinecite{Jong2}
for the calculation of the shot noise in two-terminal conductors.
\bibitem{Levinson} Joule heat in two-terminal diffusive
conductors was studied by
M.\ Rokni and Y.\ Levinson,
Phys.\ Rev.\ B{\bf 52}, 1882 (1995).
\bibitem{Gram}
Th.\ Gramespacher and M.\ B\"uttiker,
cond-mat/9805295.
\bibitem{comparison}
In this form $\Pi$ together with (\ref{MSD})
can be compared to the general result
($1.16$) of Ref.\ \onlinecite{Buttiker2}. We get
$(e^2/h)Tr[{\bf A}_{kl}(n){\bf A}_{lk}(m)]=2\int d{\bf r}
\nabla\phi_n\!\cdot\!\hat\sigma \nabla\phi_m\phi_k\phi_l$,
where ${\bf A}_{kl}(n)$ are
scattering matrices containing all information about disorder and the
shape of the conductor. The proof of this formula can be considered
as a proof of the equivalence of the classical and the quantum approach
to the calculation of the current noise spectrum in diffusive
conductors with purely elastic scattering.
However, we leave this as an open  problem .
\bibitem{comment1}
Our  $S_{24}$ differs from the one obtained in Ref.\ \onlinecite{Blanter}.
\bibitem{Brown}
R.\ Hanbury Brown and R.\ Q.\ Twiss, 
Nature (London) {\bf 177}, 27 (1956).
\bibitem{Tarucha}
R.\ C.\ Liu, B.\ Odom, Y.\ Yamamoto, and S.\ Tarucha,
Nature {\bf 391}, 263 (1998).
\bibitem{comment2}
Our universality proof
generalizes the discussion of Ref.\ \onlinecite{Jong2},
which considers a quasi-one-dimensional wire.
\bibitem{comment3}
In the case of hot electrons the expressions for $S_{nm}$
which are analogous to (\ref{MSD2}) are rather lengthy 
and we shall omit them here.
\end{references}
\end{document}